# Enhancement of Real Time EPICS IOC PV Management for Data Archiving System

**Jae-Ha Kim**

Korea Multi-purpose Accelerator Complex, Korea Atomic Energy Research Institute, Gyeongju, Korea

For operating a 100MeV linear proton accelerator, the major driving values and experimental data need to be archived. According to the experimental conditions, different data are required. It is necessary to implement functions that can add new data and delete data in real time. In an experimental physics and industrial control system (EPICS) input output controller (IOC), the process variable (PV) values are matched with the driving values and data. The PV values are archived in text file format using the Channel Archiver. There is no need to create a database (DB) server, just need a large hard disk. Through the web, the archived data can be loaded and new PV values can be archived without stopping the archive engine. Implementation of a Data Archiving System with Channel Archiver is presented and some preliminary results are reported.

# I. INTRODUCTION

In the KOrea Multi-purpose Accelerator Complex (KOMAC) for operating a 100MeV linear proton accelerator, more than 4000 machine parameters and statuses are used. They need to record to be inspected and improve the linac operation of the KOMAC. According to the experimental conditions, new machine parameters are required to archive and some that are no related should be deleted. For example, RF power values have been archived per 10 seconds. But in experiment, data, archived per a second is needed. Reconfiguration of the archive engine takes 10 minute at least and after experiment that process are needed. It is necessary to implement functions that can add new data and delete data in real time without reconfiguration. When new data are required to archive in the existing data archiving system configured using a My structured query language (MySQL) database and control system studio (CSS), the archive engine should be stopped to reconfigure the archive engine. Therefore, we need an additional data archiving system which has the functions that can add data and can be accessible through Web browser. The existing data archiving system is used to archive data, need to be archive over a long period of time. The data, necessary for experiment are archived using the additional archiving system. The additional data archiving system is configured with the Experimental Physics and Industrial Control System (EPICS) base, extensions, archive viewer and channel archive on LINUX [2]. Through the channel archiver new data are added on the web page. The Archive viewer, a user interface tool can plot written data and save data in text format, if clients are connected to network.

## II. DATA ARCHIVING

The EPICS mechanism was chosen to manage the KOMAC proton linear accelerator. A data archive system based on EPICS is needed to archive the machine parameters and data. Therefore the CSS which is related with EPICS was chosen for KOMAC. An archive engine, a CSS tool can takes PV value from EPICS IOC and store data in a database. An archive engine is configured using an XML file showing what to archive and how to do so. JDBC libraries for a MySQL database are included in the CSS archive system. A relational database (RDB) with MySQL was created for the KOMAC database. Figure 1 shows the logical data archiving system of KOMAC. The databases are redundant using MCCS. RDB1 is the main database where the archive engine usually writes data in usual. RDB2 is synchronized with RDB1. RDB2 checks the heartbeat of RDB1 every second. If RDB1 causes an error, MCCS recognizes error signal and causes interrupt. RDB2 is to replace the RDB1. RDB2 takes all data from RDB1 and becomes the main database.

The additional data archiving system is configured to the EPICS base, extensions, archive viewer and channel archiver. The archive engine in the channel archiver is configured with an XML file. The data are written to a hard disk in index and data file formats. When the archive engine is running, the archive engine will create a built-in web server. Clients can check the engine status and configuration. On the "Config" page of the web server, data can be added without stopping the archive engine. Daily restart and monitoring engine conditions are available using the Archive Daemon configuration. By running the network data server, other machine can read and access to archived data using a hypertext transfer protocol (HTTP) web server configuration [3]. The additional data archiving system is shown Fig. 2. By using an archive viewer, data can be dealt with, such as plotting data and saving data.

## III. IMPLEMENTATION

We made a test environment that consists of EPICS IOC, the data archiving system and an EPICS client on a network for testing the data archiving system. One hundred process variables (PVs) matched with the data, were created in the EPICS IOC. EPICS base R3.14.10, extensions, channel archiver, and archive viewer were installed on Centos 6.3, a Linux OS. The archive engine was configured to store 100 PVs every second. By configuring the Archive daemon and using a crontab command, data are written and updated every day. The directory, named as date of day is and data are archived in the directory using Archive daemon. Crontab command executes Archive daemon every midnight. The HTTP web server was configured to recognize the data storage path and archive data server in common gateway interface (CGI) format. The archive viewer is based on JAVA and JDK1.7 was installed. The web page of the archive data server was configured to run the archive viewer by clicking on the "Archive" text. Through the "Config" page of the engine web server, 1 additional dataset was added to test the additional functions.

The client PC needs to install a JRE or JDK to access archived data through the archive viewer. By typing in the internet protocol (IP) of the data archiving system, the webpage of the data archiving system is accessible.

## IV. CONCLUSION

PVs were taken from the data and the channel archiver scans the PV values every second with channel access. Data were stored in a specified directory. The channel archiver archived data for 3 days. The data were archived at 220 MB per day. If the number of PVs reaches about 4000, the archived data capacity will be 8.8 GB per day. A 1 TB hard disk will be used for about 120 days. The archive viewer is shown in Fig. 3. Through the archive viewer, we can load the required data, plot the data and export data in a spreadsheet and much more.

The new data archiving system of KOMAC is shown in Fig4. The major parameters and data for a long observation period are stored in the data archiving system based on CSS. Some parameters and data that are required during the experiment are recorded by the channel archiver. We can reduce the loss of data and manage the data more efficiently. To apply the new data archiving system to the KOMAC data archiving system, a central processing unit (CPU) load test with more than 4000 PVs is needed.

## ACKNOWLEDGMENT

This work is supported by the Ministry of Education, Science and Technology of the Korean Government.

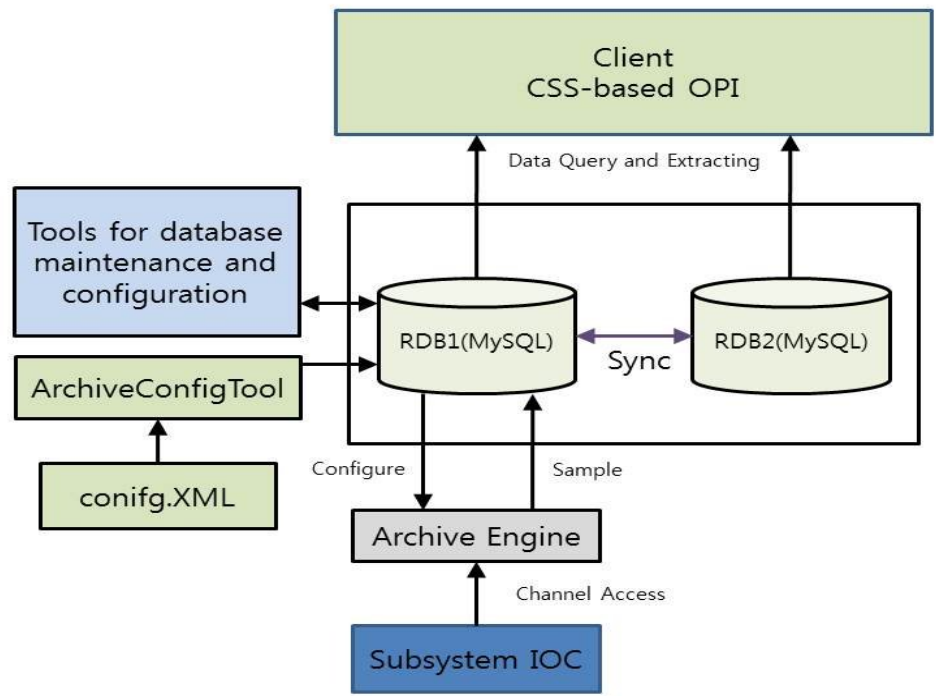

Fig. 1. Software architecture of the KOMAC EPICS data archive system.

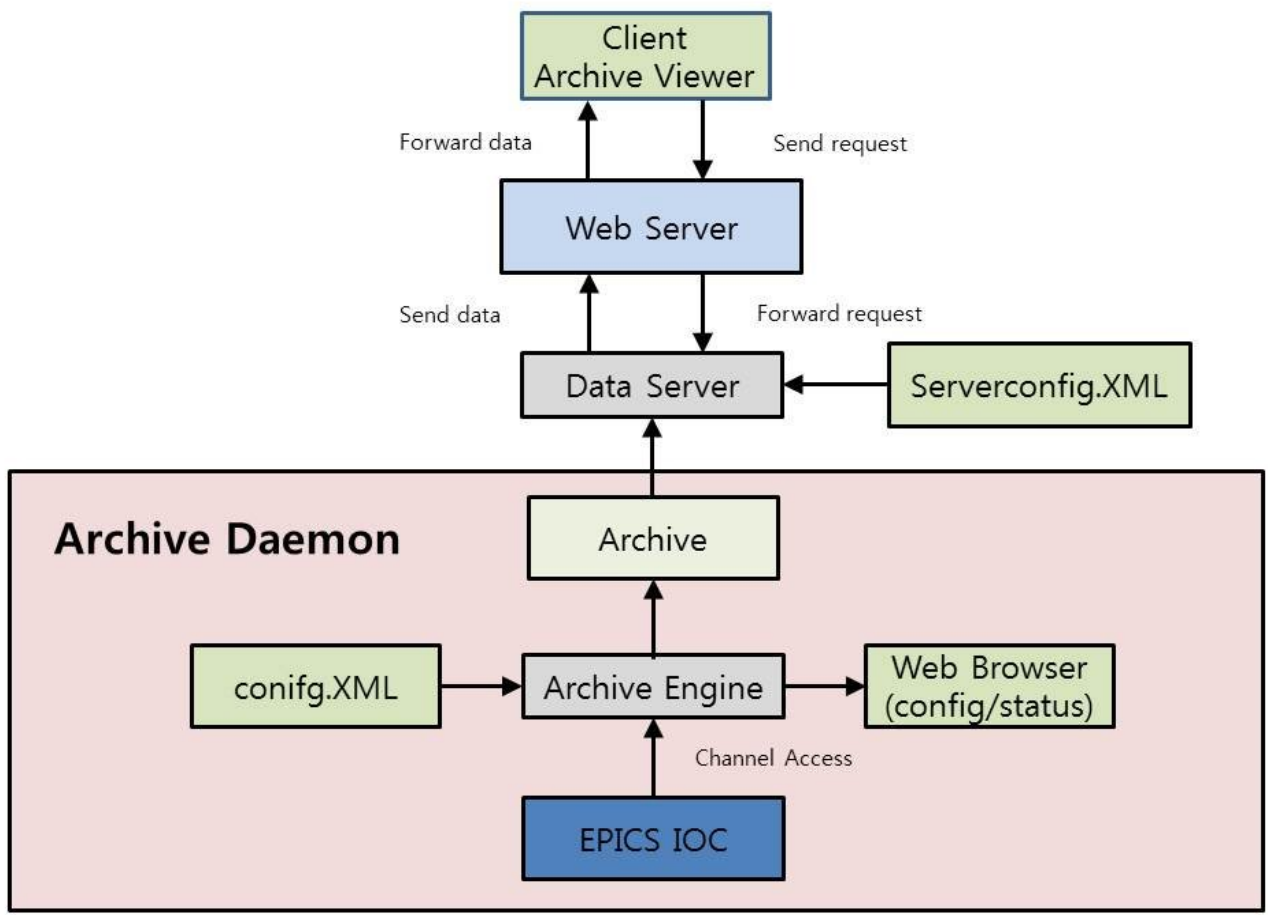

Fig. 2. Software architecture of the EPICS data archive system based on channel archiver.

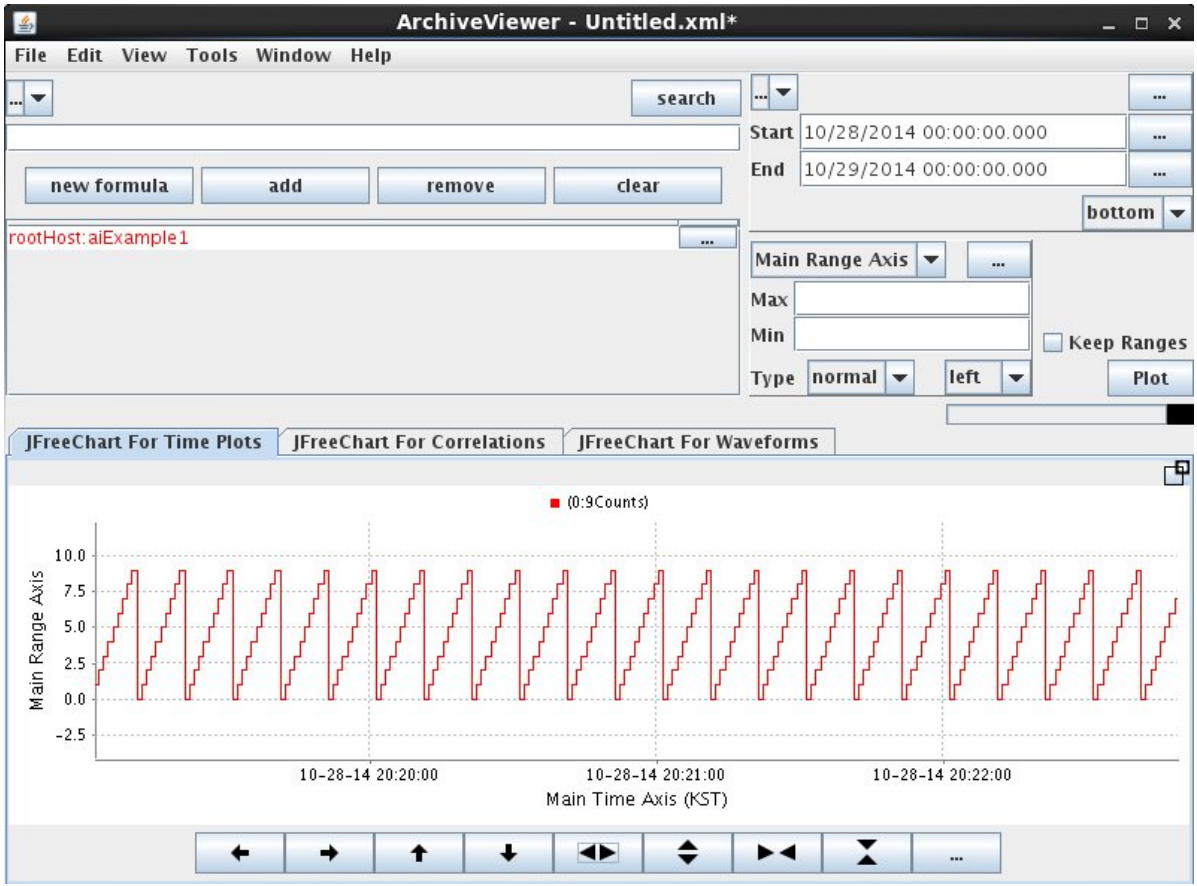

Fig. 3. GUI of archive viewer.

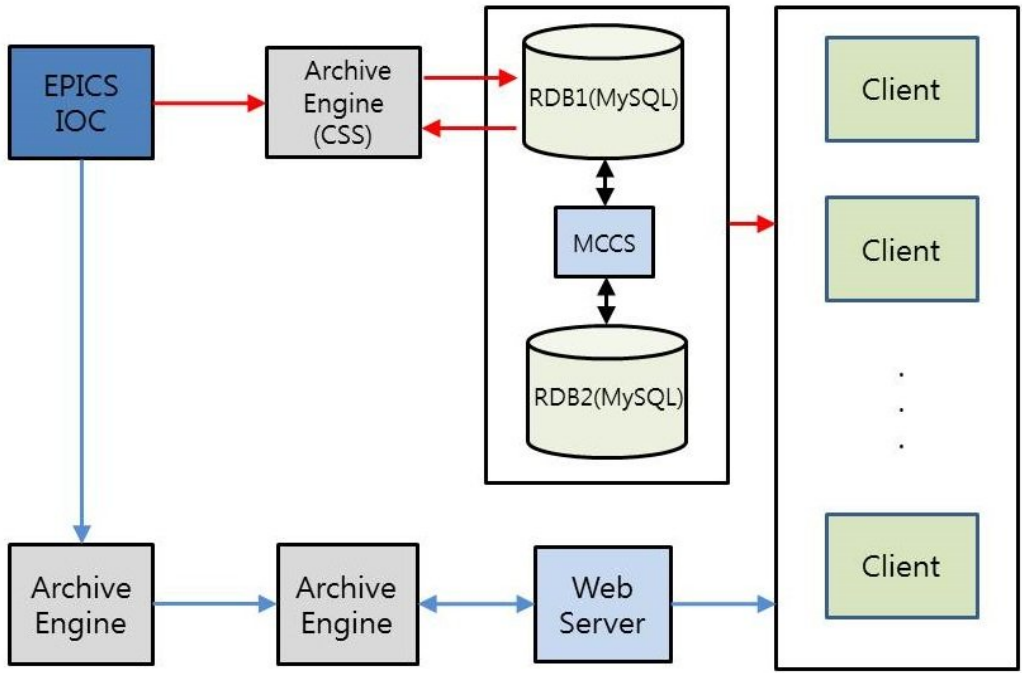

Fig. 4. Software architecture of Data archiving system.

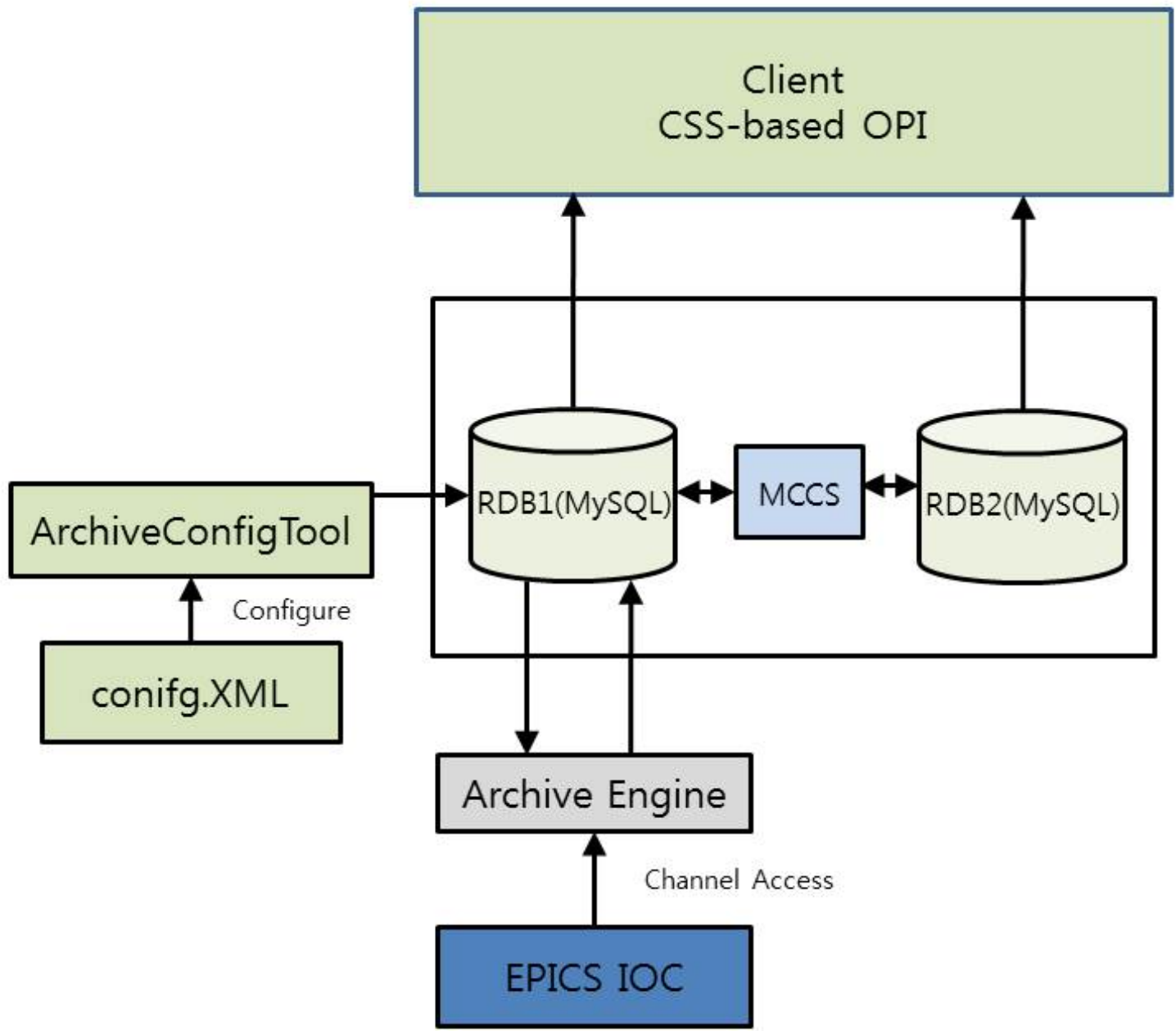

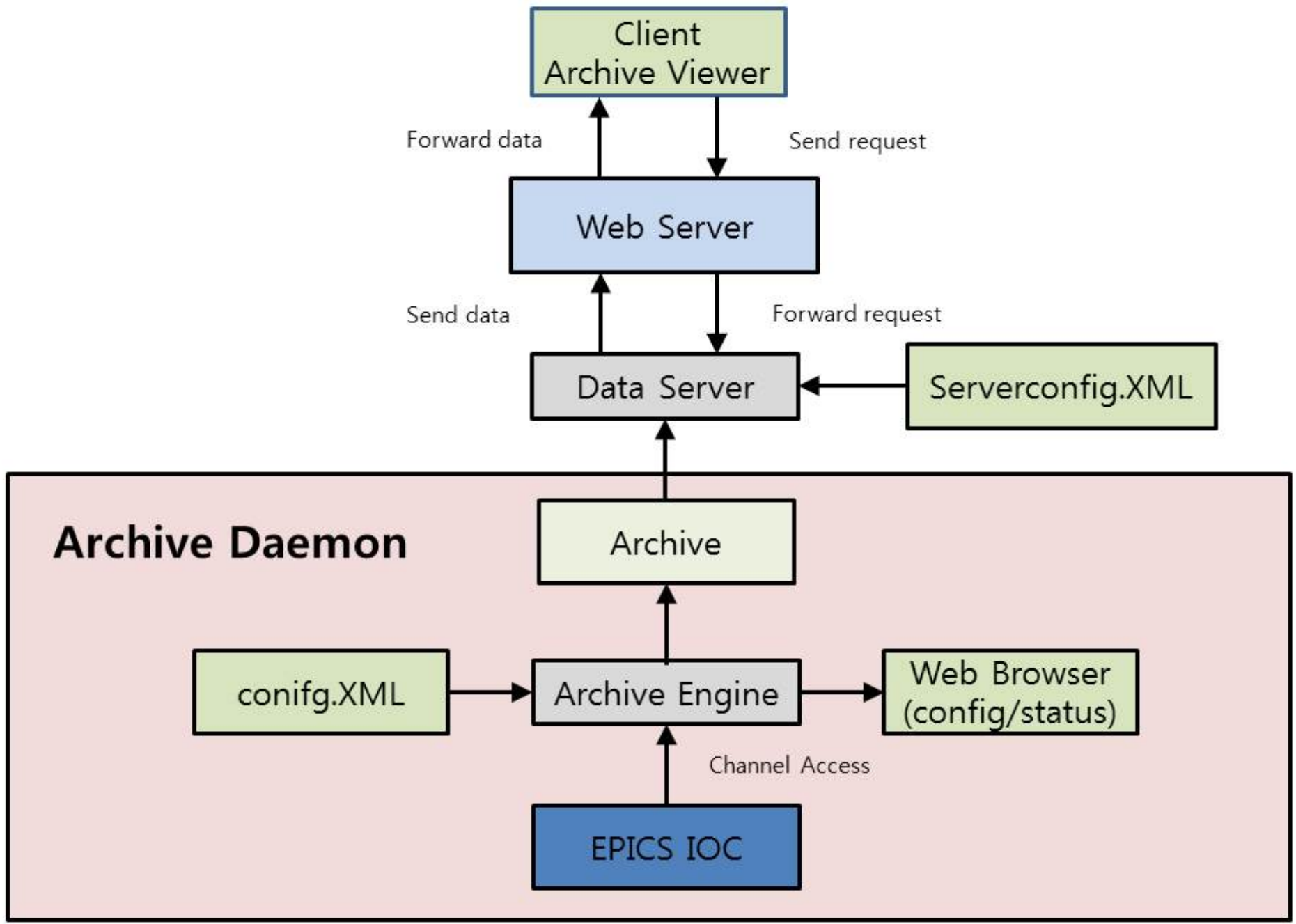

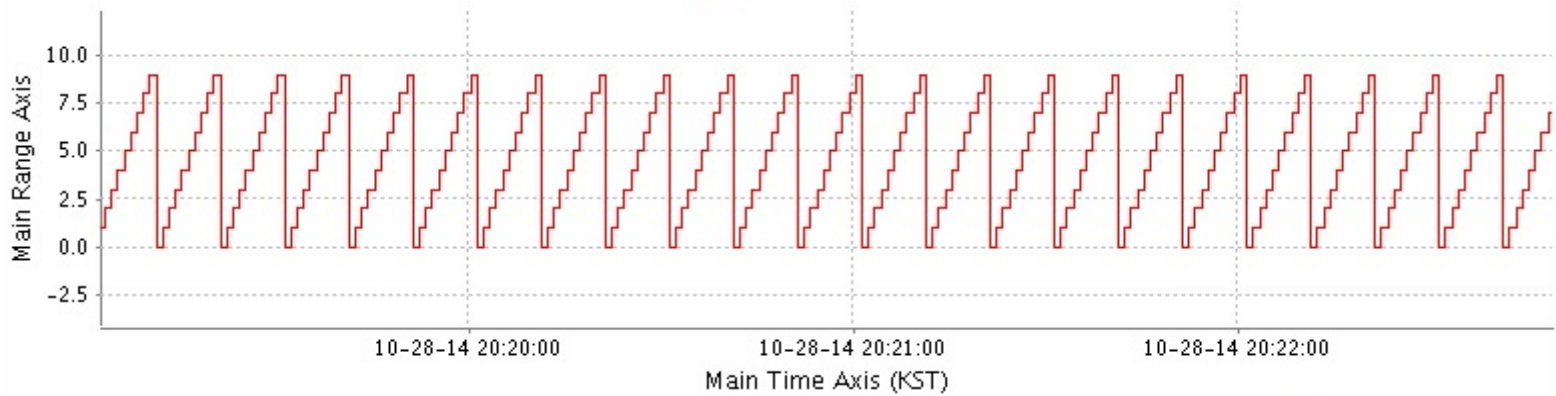

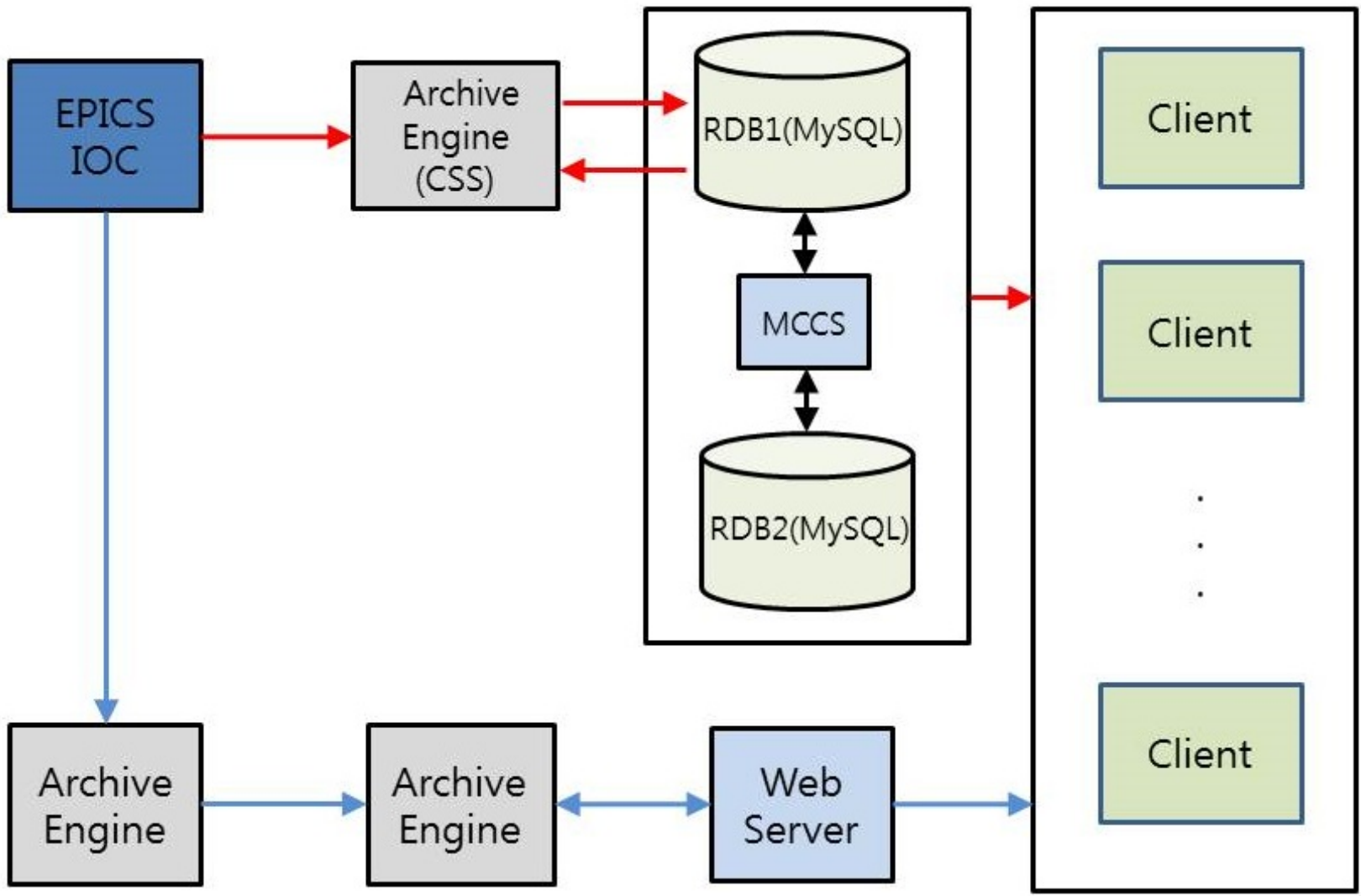